\documentclass[pdflatex,sn-mathphys,Numbered]{sn-jnl}

\usepackage{graphicx}%
\usepackage{amsfonts}%
\usepackage{xcolor}%
\usepackage{manyfoot}%
\usepackage{physics}
\usepackage[short]{optidef}

\begin{document}

\title[classical discord]{Classical Optical Analogue of Quantum Discord}

\author*[1]{\fnm{Jacob M.} \sur{Leamer}}\nomail
\equalcont{These authors contributed equally to this work.}

\author[1]{\fnm{Wenlei} \sur{Zhang}}\nomail
\equalcont{These authors contributed equally to this work.}

\author[1]{\fnm{Nicholas J.} \sur{Savino}}\nomail
\equalcont{These authors contributed equally to this work.}

\author[2]{\fnm{Ravi K.} \sur{Saripalli}}\nomail

\author[3]{\fnm{Sanjaya} \sur{Lohani}}\nomail

\author[1]{\fnm{Ryan T.} \sur{Glasser}}\email{rglasser@tulane.edu}

\author[1]{\fnm{Denys I.} \sur{Bondar}}\email{dbondar@tulane.edu}

\affil*[1]{Department of Physics and Engineering Physics, Tulane University, New Orleans, Louisiana 70118, USA}

\affil[2]{Directed Energy Research Center, Technology Innovation Institute, Abu Dhabi, UAE}

\affil[3]{Department of Electrical and Computer Engineering, University of Illinois - Chicago, Chicago, Illinois 60607, USA}

\abstract{Quantum discord has been shown to be a resource for quantum advantage in addition to quantum entanglement. While many experiments have demonstrated classical analogies of entanglement, none have done so for discord. We present a proof-of-concept demonstration for creating a classical analogue of quantum discord using classical light that takes advantage of the analogy between the state of two qubits and the spatial modes of a Laguerre-Gauss beam. We demonstrate the validity of this approach by comparing the intensity profiles of theoretical simulations to experimental results for different values of discord. Such a classical analogue of quantum discord may provide further insight in understanding and development of quantum information technologies that make use of discord.}

\maketitle

\section{Introduction}\label{sec:introduction}
The possibility of exploiting the properties of quantum systems to achieve information processing capabilities that surpass current classical approaches has motivated much interest in the field of quantum information over the last several decades \cite{Quantumsupremacy}.  This has resulted in the development of methods for quantum key distribution \cite{bb84,experimental_quantum_crypt}, quantum metrology and sensing \cite{PhysRevLett.124.230504,Quantum-EnhancedMeasurements,Quantummetrology,Gravitational-WaveDetection}, and quantum computation \cite{Feynman1982-FEYSPW,Quantumsupremacy} to name a few.  Many of these quantum information approaches rely upon the fact that quantum systems can exhibit higher levels of correlation than classical systems, such as those found in entangled quantum states. In addition to quantum entanglement, quantum discord has also been shown to be a resource for quantum advantage \cite{knill_power_1998, datta_quantum_2008, lanyon_2008}. A previous demonstration has utilized non-zero quantum discord of two-photon states to mimic quantum teleportation with a pseudo-thermal light source \cite{chen_quantum_2021}. Fundamentally, quantum discord serves as a measure of the non-classical correlations that persist even in the absence of pure quantum states \cite{ollivier_quantum_2001}. Non-zero discord represents correlations due to non-commutativity rather than non-separability \cite{luo_quantum_2008, chille_2015}.

Unfortunately, it can be difficult to realize quantum technologies due to the fragility of quantum systems. Thus, it is often beneficial to consider an analogous classical system in which it is easier to design and run experiments. Many such analogies between quantum mechanics and classical optics have long been established and are instrumental in developing our understanding of ``quantumness'' as well as leading to new optical technologies \cite{RevModPhys.17.195, PhysRevE.48.632, 1975ApPhy...6..131H, qian_shifting_2015, dragoman_2004, eberly_quantum_2016, korolkova_quantum_2019,tzanakis_discovering_1998}.  

A particularly promising classical-quantum analogy is the notion of ``classical entanglement,'' which describes classical systems with the correlations between the degrees of freedom of a classical light beam that mathematically resemble quantum entanglement\cite{korolkova_quantum_2019, liu_2019}.  While this classical notion of entanglement is an inherently local phenomenon, some works have suggested that non-locality is not required for many quantum information tasks \cite{borges_2010}.  In fact, this ``classical entanglement'' has been observed \cite{qian_entanglement_2011, spreeuw_classical_1998, mclaren_entangled_2012, aiello_quantumlike_2015, mclaren_measuring_2015, souza_2007, lee_2002, kondrad_2019, korolkova_quantum_2019} and used to implement classical analogues of quantum information algorithms \cite{sun_non-local_2015, Stoklasa_2015, goyal_2013, cardano_2015, de_oliveira_2005, perez_2018}.

To the best of our knowledge, there have been no attempts to extend these analogies to consider systems that display quantum correlations that are not entanglement. In this paper, we present a proof-of-concept demonstration for creating a classical analogue of quantum discord using classical optics. Our method uses the spatial modes of the light as an analogy to the levels of a simulated quantum system, which is described by the Bell diagonal states. This system is one of the few for which analytical results of the quantum discord are obtained \cite{luo_quantum_2008, zhou_entanglement_2013}.

The rest of the paper is organized as follows: In Sec.~\ref{sec:discord}, we recall the definition of quantum discord and apply it to the density matrix of two qubit systems. In Sec.~\ref{sec:analogy}, a classical optical analogy to quantum discord is developed using the Laguerre-Gauss spatial modes and polarization of light. In Sec.~\ref{sec:expmt}, we describe the experimental setup used to test the analytical results. The comparisons between experimental and theoretical results are given in Sec.~\ref{sec:results}. 

\section{Quantum Discord}\label{sec:discord}
Quantum discord $\mathcal{D}$ is defined as the difference between the quantum mutual information and the total classical correlations of a density matrix $\hat\rho$ \cite{ollivier_quantum_2001, Henderson_2001}.  A value of $\mathcal{D} > 0$ indicates that $\hat\rho$ contains non-classical correlations.  For many $\hat\rho$, obtaining an analytical expression for $\mathcal{D}$ is infeasible.  A notable exception is the density matrix of two qubit systems given by
\begin{align}
    \hat\rho_{BD} =& \lambda_0\ketbra{\psi^+}+\lambda_1\ketbra{\phi^+} \notag \\ &+\lambda_2\ketbra{\psi^-}+\lambda_3\ketbra{\phi^-}, \label{bd_rho}
\end{align}
where $\lambda_i$ are the eigenvalues of $\hat\rho_{BD}$ and
\begin{align}
    \ket{\psi^\pm} &= \frac{1}{\sqrt{2}}\left(\ket{00} \pm \ket{11}\right), \label{bd_psi} \\
    \ket{\phi^\pm} &= \frac{1}{\sqrt{2}}\left(\ket{01} \pm \ket{10}\right),
    \label{bd_phi}
\end{align}
are the Bell diagonal states \cite{luo_quantum_2008, zhou_entanglement_2013}.  Because analytical results exist for the discord of $\hat\rho_{BD}$, we choose to consider these states when developing our classical optical analogy for discord.  Another form of $\hat\rho_{BD}$ is 
\begin{equation}
    \hat\rho_{BD} = \frac{1}{4}(I + \sum_{i=1}^3 r_i\sigma_i\otimes\sigma_i),
\end{equation}
where $I$ is the identity matrix, $\sigma_i$ are the Pauli matrices, and $r_i$ are coefficients related to the eigenvalues by
\begin{align}
    \lambda_0 &= (1-r_1-r_2-r_3)/4, \label{l0}\\ 
    \lambda_1 &= (1-r_1+r_2+r_3)/4, \label{l1}\\ 
    \lambda_2 &= (1+r_1-r_2+r_3)/4, \label{l2}\\ 
    \lambda_3 &= (1+r_1+r_2-r_3)/4. \label{l3}
\end{align}
The quantum discord of $\hat\rho_{BD}$ is given by \cite{luo_quantum_2008, zhou_entanglement_2013}
\begin{align}
    \mathcal{D} &= 2 + \sum_{i=0}^3 \lambda_i \log_2 \lambda_i - \notag \\ & \frac{1-r}{2}\log_2(1-r) - \frac{1+r}{2}\log_2(1+r), \label{bell_state_discord}
\end{align}
with $r = \max\{|r_1|, |r_2|, |r_3|\}$.
\par
\section{Classical Quantum Analogy}\label{sec:analogy}

We begin constructing our classical analogy to quantum discord by considering a monochromatic electric field $\vec{E}$ of the form
\begin{equation}
    \vec{E}(\vec{r}_\perp, t) = e^{i\omega t}\sum_{l, k = 0}^\infty\sum_{j=H, V}c_{lkj}G_{lk}(\vec{r}_\perp)\hat{S}_j, \label{general}
\end{equation}
where $\vec{r}_\perp \in \mathbb{R}^2$ is the position perpendicular to the direction of propagation, $\omega$ is the frequency, $c_{lkj}$ is a complex coefficient, $G_{lk}(\vec{r}_\perp)$ describes the spatial components of the field, and $\hat{S}_j$ is a unit vector that denotes the polarization of the field, which is either horizontal $H$ or vertical $V$.  By taking inspiration from the formalism of quantum mechanics, Eq.~\eqref{general} can be rewritten in a more suggestive form to obtain \cite{eberly_correlation_2016}
\begin{equation}
    |E) = \sum_{l, k =0}^\infty\sum_{j=H,V}c_{lkj}|G_{lk})|j), \label{classical}
\end{equation}
where the parent $(\cdot|$ and thesis $|\cdot)$ serve as classical equivalents to the bra $\bra{\cdot}$ and ket $\ket{\cdot}$ in quantum mechanics \cite{spreeuw_classical_1998}. In Eq.~\eqref{classical}, $|G_{lk})$ is defined as $( \vec{r}_\perp |G_{lk}) \equiv G_{lk}(\vec{r}_\perp)$.  We have also dropped the time dependence because it acts as a global phase in a monochromatic field. The matrix $|E)(E|$ is then the coherency matrix \cite{wolf}, which serves as an optical equivalent to the density matrix.

This approach to developing an analogy between quantum mechanics and classical optics is not novel.  The use of parenthesis as a classical version of the braket notation in quantum mechanics was first introduced by Spreeuw \cite{spreeuw_classical_1998} and the parallel between the functional and state representations of a general electric field in Eqs.~\eqref{general} and \eqref{classical} was developed by Eberly \cite{eberly_correlation_2016}. Such analogies have also seen use in a number of experiments \cite{sun_non-local_2015,Stoklasa_2015,qian_entanglement_2011}.  To extend these to the case of quantum discord, we choose a specific form of Eq.~\eqref{classical} that is mathematically equivalent to a quantum state with non-zero discord. We choose to create an optical analogy based on analytical results for the discord of $\hat\rho_{BD}$, which are shown in Eq.~\eqref{bell_state_discord}. Inspired by Ref.~\cite{liu_2019}, we use the Laguerre-Gauss beams \cite{goubau_lg} to arrive at 
\begin{align}
    |\psi^\pm) &= \frac{1}{\sqrt{2}}[|LG_{00}) \pm |LG_{11})], \\
    |\phi^\pm) &= \frac{1}{\sqrt{2}}[|LG_{01}) \pm |LG_{10})].
\end{align}
In this case, the spatial mode indices serve as an analogy to the qubits in Eqs.~\eqref{bd_psi} and \eqref{bd_phi}. Since there are still states with non-zero discord when two of the eigenvalues in Eq.~\eqref{bd_rho} are 0, we can simplify the analogy further by considering only $|\psi^+)$ and $|\phi^+)$.  In this case, $\hat\rho_{BD}$ is equivalent to the coherency matrix
\begin{equation}
    \hat\rho_{LG} = \lambda_0^{LG}|\psi^+)(\psi^+| + \lambda_1^{LG}|\phi^+)(\phi^+|,
\end{equation}
where $\lambda_i^{LG} = I_i/I_T$ is the ratio of the intensity of a state $I_i$ ($i=0$ for $|\psi^+)$ and $i=1$ for $|\phi^+)$) to the total intensity, $I_T=I_0+I_1$, and serve as optical eigenvalues. Note that $\hat\rho_{LG}$ is a mixture of two beams - one in state $|\psi^+)$ and the other in state $|\phi^+)$. To obtain such a state, we require an auxiliary degree of freedom. Here, we choose polarization for this role. We can then proceed by polarizing the beam with state $|\psi^+)$ into $H$-polarization and the beam with state $|\phi^+)$ into $V$-polarization and combining them on a beamsplitter, which yields the state,

\begin{equation}
    |E) = \sqrt{\lambda_0^{LG}}|\psi^+)|H) + \sqrt{\lambda_1^{LG}}|\phi^+)|V). \label{incoh_sum}
\end{equation}
Finally, we can take the partial trace of $|E)(E|$ over the polarization space to obtain
\begin{equation}
    \hat\rho_{LG} = \Tr_{|j)}\{|E)(E|\}.
\end{equation}  
Experimentally, the partial trace over polarization can be achieved by capturing the intensity profile with a polarization insensitive camera \cite{Barberena:15}.  To obtain the discord for a given $\hat\rho_{LG}$, we first solve for $r_1, r_2,$ and $r_3$ using Eqs.~\eqref{l0}-\eqref{l3} with $\lambda_0=\lambda_0^{LG}$, $\lambda_1=\lambda_1^{LG}$, and $\lambda_2=\lambda_3=0$, then the value of the discord can be calculated using Eq.~\eqref{bell_state_discord}.

\section{Experiment}\label{sec:expmt}

\begin{figure}[htbp]
    \centering
    \includegraphics[width = .8\linewidth]{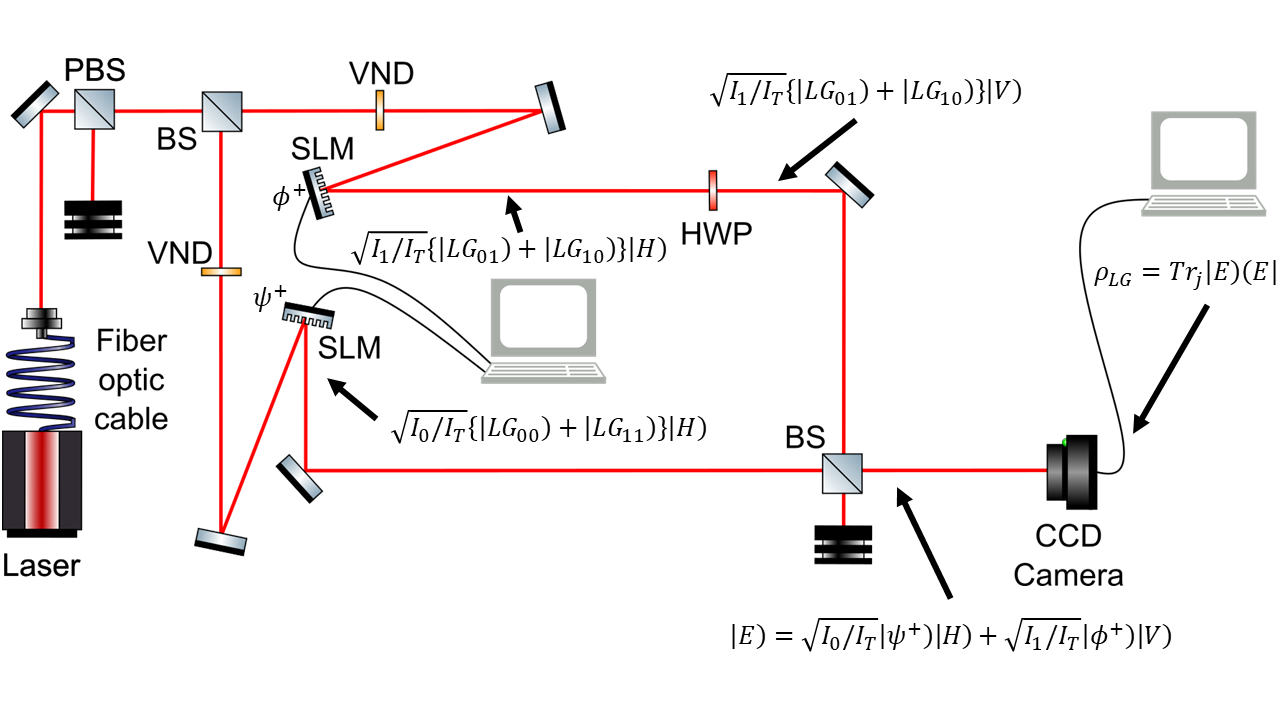}
    \caption{MZI setup for generation of the state $|E) = \sqrt{\frac{I_0}{I_T}}|\psi^+)|H) + \sqrt{\frac{I_1}{I_T}}|\phi^+)|V)$. Abbreviations: $\lambda /2$ = half-waveplate, SLM = spatial light modulator, PBS = polarizing beamsplitter, BS = 50:50 beamsplitter, VND = variable neutral density filter, CCD = charge coupled device.}
    \label{fig:exp setup}
\end{figure}

We use a setup based on the Mach Zehnder interferometer (MZI) for the generation of the required state, $|E)$, as shown in Fig.\ref{fig:exp setup}. A laser beam with 795nm wavelength emitted from an amplified diode laser is used for our experiments. The total intensity ($I_T$) into the MZI is controlled using a half-wave plate and polarizing beamsplitter. A 50:50 beamsplitter splits the beam of horizontally polarized light into two arms. Each arm of the MZI contains a variable neutral density filter, such that we control the ratio of intensities. The spatial light modulators in the left and right arms project the states $|\psi^+) = |LG_{00}) + |LG_{11})$ and $|\phi^+) = |LG_{01}) + |LG_{10})$ respectively. Another half-wave plate is placed in the right arm to change the initial horizontal polarization to vertical polarization for that state. The two states are combined using a 50:50 beamsplitter, thus generating the required state $|E) = \sqrt{\lambda_0^{LG}}|\psi^+)|H) + \sqrt{\lambda_1^{LG}}|\phi^+)|V)$. A CCD camera performs the partial trace of $|E)(E|$ over polarization and is used to capture the two dimensional intensity profile $(\vec{r}_\perp|\hat\rho_{LG}|\vec{r}_\perp)$.  The images of the individual states, $|(\vec{r}_\perp|\psi^+)|^2$ and $|(\vec{r}_\perp|\phi^+)|^2$, can be obtained by blocking the required arm of the MZI.
\begin{figure}[h!]
    \centering
    \includegraphics[width = .8\linewidth]{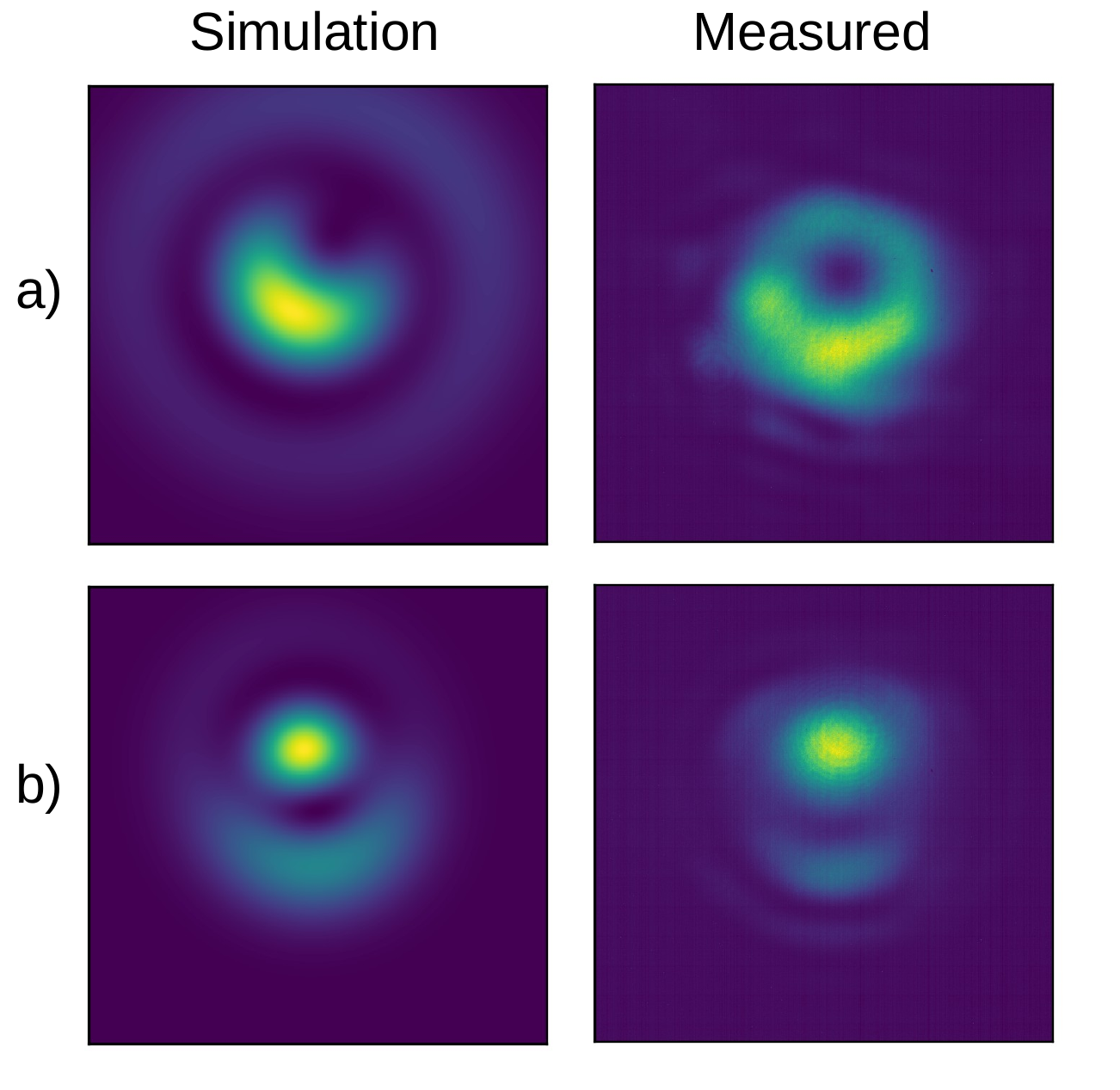}
    \caption{Simulated and experimentally measured intensity profiles a) $|(\vec{r}_\perp|\psi^+)|^2$ and b) $|(\vec{r}_\perp|\phi^+)|^2$.  The simulated intensity profiles are obtained using the analytical form of the Laguerre-Gauss modes to construct the states a) $|\psi^+)$ and b) $|\phi^+)$.  The experimentally measured intensity profiles are obtained by blocking the appropriate arm of the MZI setup as described in Section~\ref{sec:expmt}.  The fact that the simulated and experimentally determined intensity profiles display the same prominent features in both cases confirms that our setup is able to correctly generate the required states.}
    \label{fig:phi&psi}
\end{figure}

To test the validity of our classical optical analogy of quantum discord, we compare the intensity profiles of our experimental results to the images obtained using the analytical forms for the Laguerre-Gauss modes \cite{goubau_lg}.  We start by looking at the individual states before they are combined.  The intensity profiles for the theoretical and experimental results are displayed in Fig.~\ref{fig:phi&psi}.  For both $|\psi^+)$ and $|\phi^+)$, the experimental images are very similar to the simulated images, which indicates that our experiment is able to create the individual states.  The constrained spatial resolution of the CCD results in subtle divergences between the simulated and experimental images. Minor misalignment of the spatial light modulators also can contribute slightly to the experimental error.

\begin{figure}[h!]
    \centering
    \includegraphics[width = .9\linewidth]{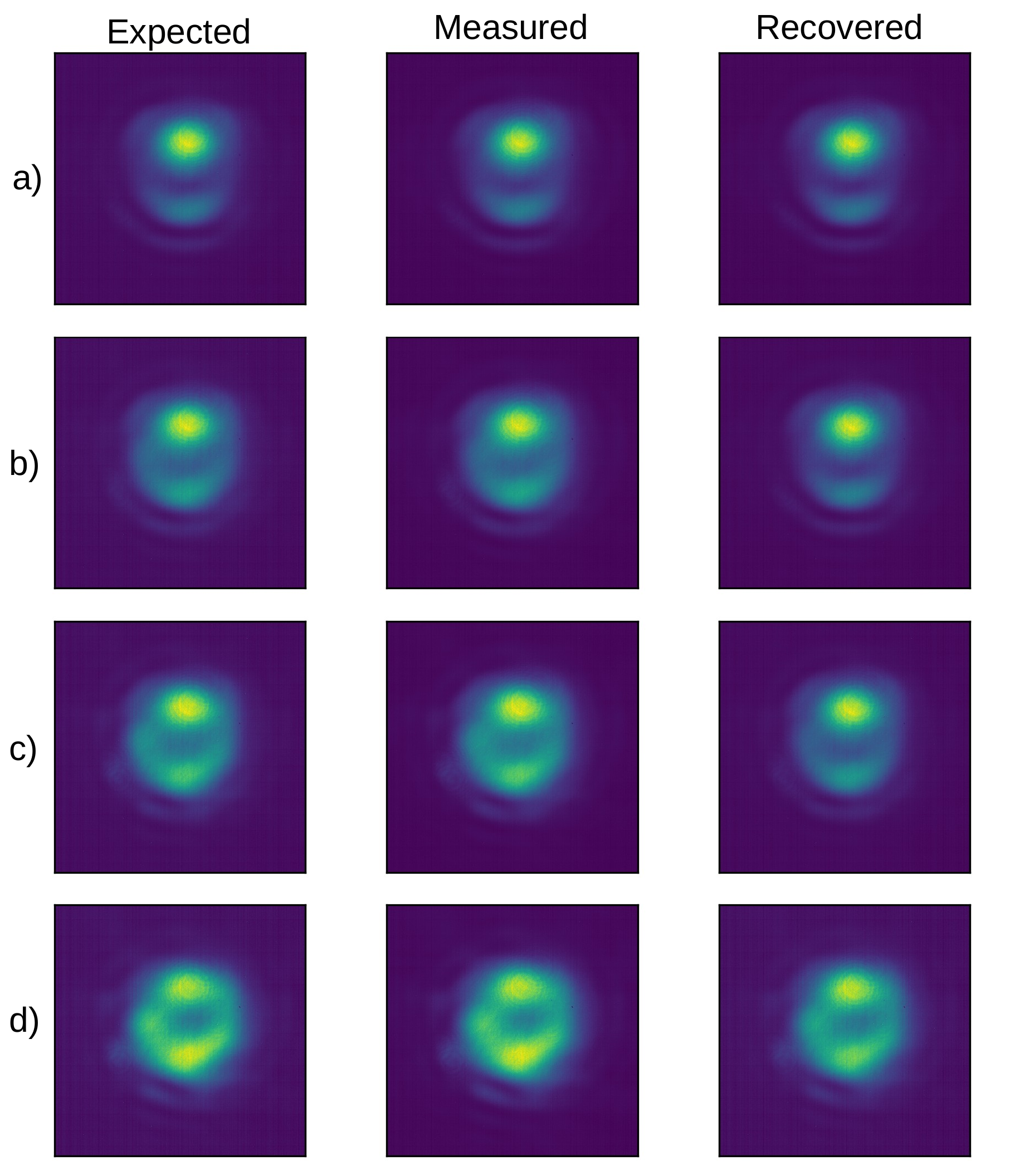}
    \caption{Intensity profiles corresponding to the state $|E) = \sqrt{\lambda_0^{LG}}|\psi^+)|H) + \sqrt{\lambda_1^{LG}}|\phi^+)|V)$ for $\lambda_0^{LG} = $ a) 0.17, b) 0.38, c) 0.49, and d) 0.64. The experimental intensity profiles $(\vec{r}_\perp|\hat\rho_{LG}|\vec{r}_\perp)$ in the middle column are measured by the CCD as shown in Fig.~\ref{fig:exp setup}. The expected intensity profiles in the left column are calculated using Eq.~\eqref{constructed}. The recovered intensity profiles in the right column are calculated using Eq.~\eqref{recovered}.}
    \label{fig:E}
\end{figure}

To obtain the specific state with a certain discord value with our experimental setup, we first solve for $\lambda_0$ and $\lambda_1$ using Eqs.~\eqref{l0}-\eqref{bell_state_discord} with $\lambda_2=\lambda_3=0$ and the required value of $\mathcal{D}$, then we adjust the variable neutral density filters in the setup such that $\lambda_0^{LG}=\lambda_0$ and $\lambda_1^{LG}=\lambda_1$ before combining the two beams. We then compare the measured intensity profiles of the combined beam (middle column of Fig.~\ref{fig:E}) to the expected intensity profiles (left column of Fig.~\ref{fig:E}), which are calculated from

\begin{gather}
    P_{\textrm{expect}} = \lambda_0^{LG}|(\vec{r}_\perp|\psi^+)|^2 + \lambda_1^{LG}|(\vec{r}_\perp|\phi^+)|^2, \label{constructed}
\end{gather}
where $|(\vec{r}_\perp|\psi^+)|^2$ and $|(\vec{r}_\perp|\phi^+)|^2$ are the measured intensity profiles of $|\psi^+)$ and $|\phi^+)$, respectively, shown in Fig.~\ref{fig:phi&psi}. In all cases, both sets of profiles display the same prominent features, which indicates that the intensity profiles of the combined beams are close to the expected ones.

\section{Results}\label{sec:results}

\begin{figure}[h!]
    \centering
    \includegraphics[width = .9\linewidth]{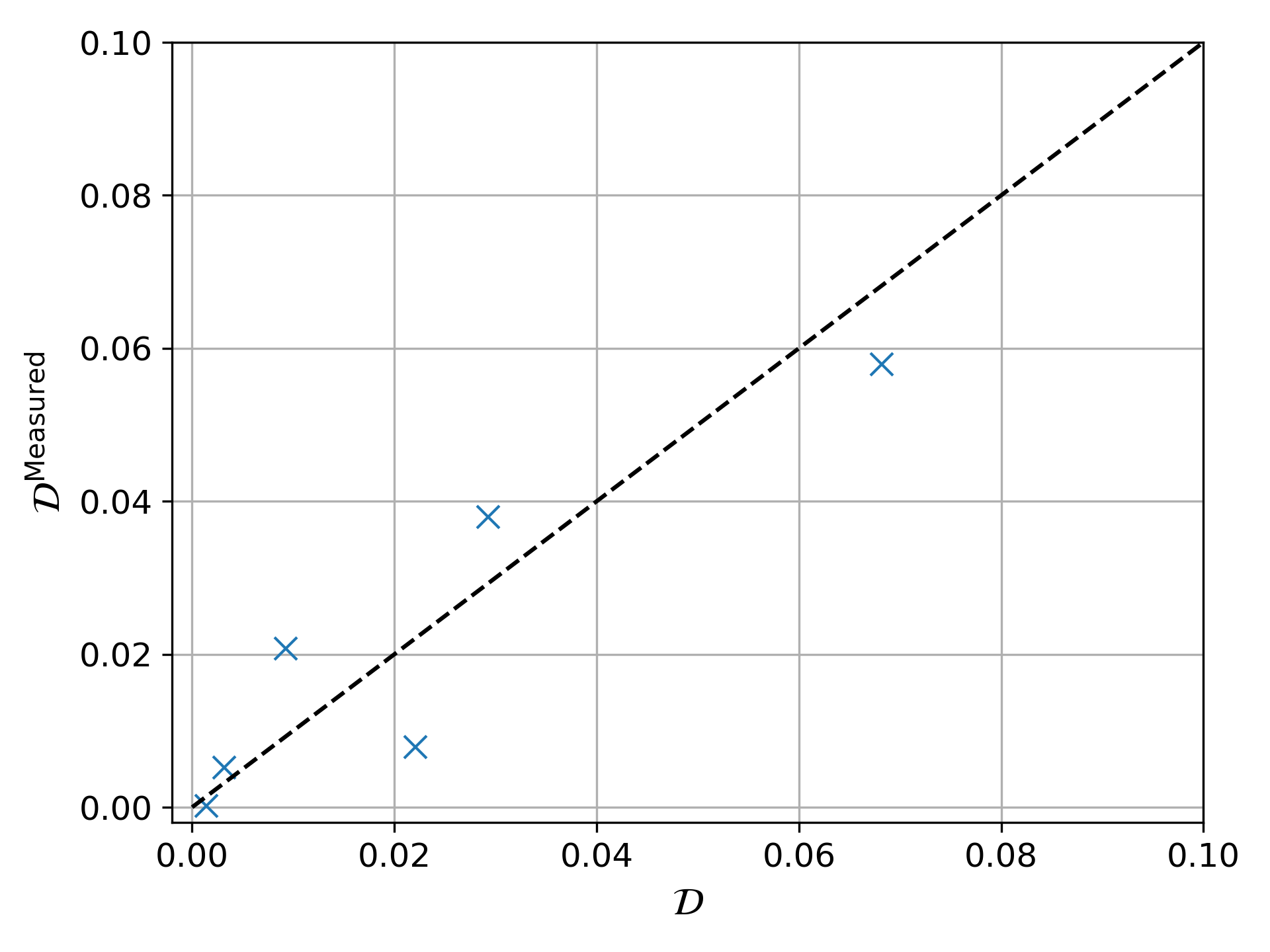}
    \caption{Comparison of measured and required values of discord. The required value of discord, $\mathcal{D}$, is chosen at the beginning of the experiment. The measured value of discord, $\mathcal{D}^\textrm{measured}$, is calculated from the recovered intensity ratios outlined in Section~\ref{sec:results}. The diagonal dashed line represents equality between the values on the horizontal and vertical axes.}
    \label{fig:measured_vs_setting}
\end{figure}

In this section, we would like to verify that the measured state at the camera has the required value of discord with the procedure presented in the previous section. Since it is not possible to experimentally determine $\lambda_0^{LG}$ and $\lambda_1^{LG}$ of the combined beam with just the intensity profile measured by the camera, we employ the following method to recover these values indirectly. We solve the optimization problem
\begin{mini}<b>
    {x}{\int d\vec{r}_\perp\big[(\vec{r}_\perp|\hat\rho_{LG}|\vec{r}_\perp) - x|(\vec{r}_\perp|\psi^+)|^2}{}{}
    \breakObjective{- (1-x)|(\vec{r}_\perp|\phi^+)|^2\big]^2}
    \addConstraint{0 \leq x \leq 1,}{}{\labelOP{opt_prob}}{}
\end{mini}
for $x$ using the experimentally measured images $(\vec{r}_\perp|\hat\rho_{LG}|\vec{r}_\perp)$ (middle column of Fig.~\ref{fig:E}) and $|(\vec{r}_\perp|\psi^+)|^2$ and $|(\vec{r}_\perp|\phi^+)|^2$ (right column of Fig.~\ref{fig:phi&psi}). Thus, the recovered intensity ratios are given by $\lambda_0^\textrm{rec}=x$ and $\lambda_1^\textrm{rec}=1-x$. To verify the validity of this method, we compare the experimentally measured intensity profiles (middle column of Fig.~\ref{fig:E}) to the recovered intensity profiles (right column of Fig.~\ref{fig:E}), which are calculated from

\begin{gather}
    P_{\textrm{rec}} = \lambda_0^\textrm{rec}|(\vec{r}_\perp|\psi^+)|^2 + \lambda_1^\textrm{rec}|(\vec{r}_\perp|\phi^+)|^2. \label{recovered}
\end{gather}
In all cases, both sets of profiles display the same prominent features, which indicates the recovered intensity ratios are close to the experiment. Finally, the measured value of discord, $\mathcal{D}^\textrm{measured}$, are calculated using Eqs.~\eqref{l0}-\eqref{bell_state_discord} with $\lambda_0=\lambda_0^\textrm{rec}$, $\lambda_1=\lambda_1^\textrm{rec}$, and $\lambda_2=\lambda_3=0$ and compared to the required value of discord, $\mathcal{D}$, chosen at the beginning of the experiment. Figure~\ref{fig:measured_vs_setting} shows the result of this comparison. In Fig.~\ref{fig:measured_vs_setting}, the diagonal dashed line represents equality between the measured and required values of discord. Our results show reasonable agreement between the measured and required values. The deviation from the diagonal line is mainly due to fluctuations of the laser and imperfections of optical components. These same technical issues also make it difficult to obtain accurate results for values of discord higher than 0.1.

\section{Conclusion}\label{sec:conclusion}
We present a method for creating a classical analogue of the quantum discord in two-qubit Bell states using classical light with spatial components described by the Laguerre-Gauss modes.  We demonstrate the validity of our setup by comparing the intensity profiles $|(\vec{r}_\perp|\psi^+)|^2$ and $|(\vec{r}_\perp|\phi^+)|^2$ with analytical solutions in Fig.~\ref{fig:phi&psi} and by comparing the combined beam's measured intensity profile $(\vec{r}_\perp|\hat\rho_{LG}|\vec{r}_\perp)$ with the expected intensity profile $P_{\textrm{expect}}$ [Eq.~\eqref{constructed}] in Fig.~\ref{fig:E}.  We also introduce a scheme for measuring the discord at the CCD that is based on solving the optimization problem~\eqref{opt_prob}.  The measured values of the discord is compared to the values of discord chosen at the beginning of the experiment in Fig.~\ref{fig:measured_vs_setting} and a reasonable agreement between is observed.  The presented method can be extended to simulate more complex systems where the quantum discord may not have analytical forms.

The analogy presented in this work can be made stronger by demonstrating that a local projective measurement has an effect on the properties of the entire system. As an example, one could show that swapping the order in which two measurements are applied changes the outcome.  For Laguerre-Gauss beams, methods of performing projective measurements have already been developed \cite{roux_2014} and experimentally used \cite{liu_2019}.  The experimental setup in Fig.~\ref{fig:exp setup} can be modified to implement these methods.

This classical analogue of quantum discord may aid the development of quantum information technologies utilizing discord, such as device-dependent quantum cryptography \cite{pirandola_quantum_2014}, quantum machine learning\cite{ghobadi_power_2019}, and remote state preparation \cite{dakic_quantum_2012}. It may also provide insights into  the boundary between classical and quantum mechanics \cite{modi_classical-quantum_2012}.

\backmatter

\bmhead{Funding} U. S. Office of Naval Research (N000141912374); Defense Advanced Research Projects Agency (D19AP00043); U.S. Army Research Office (W911NF-19-1-0377). 

\bmhead{Acknowledgments}
This material is based upon research supported by, or in part by, the U. S. Office of Naval Research under award number N000141912374. This work was also supported by the Defense Advanced Research Projects Agency (DARPA) grant number D19AP00043 under mentorship of Dr. Joseph Altepeter. D.I.B. is also supported by the U.S. Army Research Office (ARO) under grant W911NF-19-1-0377.  J. M. L. was supported by the Louisiana Board of Regents’ Graduate Fellowship Program. The views and conclusions contained in this document are those of the authors and should not be interpreted as representing the official policies, either expressed or implied, of DARPA, ONR, ARO, or the U.S. Government. The U.S. Government is authorized to reproduce and distribute reprints for Government purposes notwithstanding any copyright notation herein.

\bmhead{Disclosures} The authors declare no conflicts of interest.

\bmhead{Data availability} Data underlying the results presented in this paper are not publicly available at this time but may be obtained from the authors upon reasonable request.

\bibliography{literature}

\end{document}